\documentclass[useAMS,usenatbib]{mn2e}
\usepackage{graphicx,rotate,url,mathptmx,lscape,times, color}
\def\gtsim{\mathrel{\hbox{\rlap{\hbox{\lower4pt\hbox{$\sim$}}}\hbox{$>$}}}}
\def\lesssim{\mathrel{\hbox{\rlap{\hbox{\lower4pt\hbox{$\sim$}}}\hbox{$<$}}}}
\def\Msunpyr{M$_{\odot}\,$yr$^{-1}$}
\def\Msun{M$_{\odot}$}
\def\cm{{\rm\thinspace cm}}

\def\erg{{\rm\thinspace erg}}

\def\km{{\rm\thinspace km}}

\def\Msun{\hbox{$\rm\thinspace M_{\odot}$}}

\def\pc{{\rm\thinspace pc}}
\def\s{{\rm\thinspace s}}
\def\ps{{\rm\thinspace s^{-1}}}

\def\yr{{\rm\thinspace yr}}

\def\ergpscmps{\hbox{$\erg\cm^{-2}\s^{-1}\,$}}

\def\kmps{\hbox{$\km\ps\,$}}

\def\Msunpyr{\hbox{$\Msun\yr^{-1}\,$}}

\def\ha{\hbox{{\rm H}$\alpha$}}

\def\h0{\hbox{{\rm H}$^0$}}
\DeclareMathAlphabet{\vib}{OML}{cmm}{m}{it}
\def\z{\hbox{\it z}}

\def\NB{\hbox{$NB$}}
\def\rg{\hbox{{\rm 4C\,$+$10.48}}}
\def\nbexcess{$NB-$excess}
\def\pc{proto-cluster}
\def\RLF{$R$-LF}
\begin{document}

\title[A comparison of \pc\ and field \ha\ emitters at $z\sim2$]{H$\alpha$ emitters in $z\sim2$ \pc s: evidence for faster evolution in dense environments}
\author[Hatch]
       {\parbox[]{6.0in}
       {N.\,A.\,Hatch$^{1}$\thanks{E-mail: nina.hatch@nottingham.ac.uk}, J.\,D.\,Kurk$^2$, L.\,Pentericci$^3$, B.\,P.\,Venemans$^4$, E.\,Kuiper$^{5}$, G.\,K.\,Miley$^{5}$, H.\,J.\,A.\,R\"ottgering$^{5}$.\\
        \footnotesize
        $^1$School of Physics and Astronomy, University of Nottingham, University Park, Nottingham NG7 2RD\\
        $^2$Max-Planck-Institut f{\"u}r Astrophysik, Karl-Schwarzschild Strasse 1, D-85741 Garching, Germany\\
        $^3$INAF, Osservatorio Astronomica di Roma, Via Frascati 33, 00040 Monteporzio, Italy \\
          $^4$European Southern Observatory, Karl-Schwarzschild-Str. 2, D-85748 Garching, Germany\\
            $^5$Leiden Observatory, University of Leiden, P.B. 9513, Leiden 2300 RA, The Netherlands
    }}
 \date{}

\pubyear{}

\maketitle

\label{firstpage}
\begin{abstract}

This is a study of  \ha\ emitters in two dense galaxy \pc s surrounding radio galaxies at $z\sim2$. We show that the \pc\ surrounding MRC\,1138-262 contains $14\pm2$ times more \ha\ candidates than the average field ($9\sigma$  significance), and the $z=2.35$ radio galaxy \rg\ is surrounded by $12\pm2$ times more emitters than the field ($5\sigma$), so it is also likely to reside in a dense \pc\ environment. We compared these \ha\ emitters, situated in dense environments, to a control field sample selected from 3 separate fields forming a total area of $172$\,arcmin$^2$. 

We constructed and compared \ha\ and rest-frame $R$ continuum luminosity functions of the emitters in both environments. The star formation density is on average 13 times greater in the \pc s than the field at $z\sim2$, so the total star formation rate within the central 1.5\,Mpc of the \pc s exceeds 3000\Msunpyr. However, we found no significant difference in the shape of the \ha\ luminosity functions, implying that environment does not substantially affect the strength of the \ha\ line from strongly star forming galaxies. 

The \pc\ emitters are typically 0.8\,mag brighter in rest-frame $R$ continuum than field emitters, implying they are twice as massive as their field counterparts at the same redshift. We also show the \pc\ galaxies have lower specific star formation rates than field galaxies, meaning the emitters in the dense environments formed more of their stars earlier than the field galaxies. We conclude that galaxy growth in the early Universe was accelerated in dense environments, and that cluster galaxies differed from field galaxies even before the cluster had fully formed.

\end{abstract}
\section{Introduction}

Dense large-scale structures have been discovered in the vicinity of high redshift radio galaxies (HzRGs)  up to $z=5.2$ \citep[e.g.,][]{LeFevre1996,Pentericci2000,Kurk2000,Kurk2004a,Best2003,Overzier2006,Venemans2007,Overzier2008,Hatch2010,Galametz2010}. 
These overdense fields are some of the densest regions in the early Universe. They are generally referred to as \pc s, since: (i)  their high masses suggest they may evolve into group or cluster-sized structures \citep{Venemans2007}, (ii)  kinematic analysis shows that these structures have already begun to collapse (Kuiper et al. submitted) but are not yet fully collapsed, and (iii) the lack of extended X-ray emission expected from a shock-heated intracluster medium suggests they are not yet virialized \citep{Carilli2002}.

These \pc s comprise of concentrations of star forming galaxies, and there is little evidence of an accompanying passively evolving population. However, passive galaxies are hard to observe and spectroscopically confirm at $z>2$, so the lack of evidence does not mean passively evolving galaxies do not exist in \pc s. Recently, \citet{Gobat2011} found a red galaxy population in a $z\sim2$ cluster.

In the absence of strong feedback, galaxy formation is predicted to be more efficient in dense environments due to the abundance of surrounding gas \citep{Gunn1972}. Therefore one may expect the galaxies in dense environments to be more massive and further developed than galaxies in the field or in voids. Indeed low and intermediate redshift clusters contain more massive galaxies than the field \citep{Baldry2006,Scodeggio2009}, and their early-type members are $0.4-2$\,Gyrs older than field galaxies at the same redshift \citep{Thomas2005,vanDokkum2007,Gobat2008}. When do these differences develop and what drives them?  Are the differences between cluster and field galaxies inherent to the galaxies, driven by the initial matter distribution and abundance of gas supply, or does environment metamorphose galaxies? 

Proto-clusters provide us with a laboratory to test how environment affects galaxy formation and evolution during the formation epoch of cluster galaxies ($z>1.5$), when early-type galaxies formed the bulk of their stellar mass. Determining whether cluster galaxies differ from field galaxies before they even belonged to a cluster, tells us whether any of the difference observed today is driven by nature as apposed to nurture. 
\begin{figure}
 \begin{center}
 \includegraphics[width=1\columnwidth]{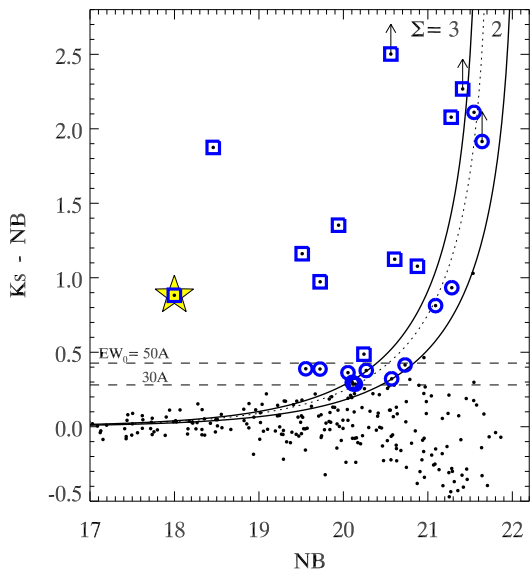}
\caption{\label{ha_select} A colour-magnitude diagram of all sources detected in the \NB\ image of \rg. The dashed lines mark the $Ks-NB$ colours equivalent to $EW_0$ of 30\AA\ and $50$\AA, whilst the solid lines mark the $Ks-NB$ colour corresponding to 2 and 3 times the combined uncertainty in the \NB\ and $Ks$ images ($\Sigma$).  \ha\ candidates were selected as sources with $EW_0>30$\AA\ and $\Sigma>2$ and are in highlighted blue. Squares mark the more-secure \ha\ candidates that correspond to the stricter criteria of $EW_0>50$\AA\ and $\Sigma>3$. The yellow star marks the radio host galaxy. The dotted line marks $\Sigma_{\rm COSMOS}>2$, the uncertainty in the shallowest control field. Only \ha\ candidates above this limit were used for comparison with the field sample.}
\end{center}
\end{figure}

To observe and interpret environmental dependences it is essential to select a clean sample of \pc\ galaxies and compare with a large field sample. Therein lies the challenge. Photometric redshifts at $z>1.5$ obtained with broad-band photometry have a limited accuracy, no more than $\Delta z/(1+z)>0.06$, which is insufficient to determine cluster membership at $z>2$ \citep{Cooper2005}. This problem is compounded further by the star-forming nature of the \pc\ galaxies, as their spectral breaks at $4000$\AA\ are weak, which yields a less accurate photometric redshift. Proto-cluster studies that use broad-band photometric redshifts are plagued by sample contamination \citep[e.g.][]{Kuiper2010,Tanaka2010b}, and any signature of environmental dependance may be washed out.

The most efficient method to obtain a clean sample of \pc\ galaxies is to select line-emitting galaxies using narrow-band imaging. Spectroscopic confirmation of line-emitting protocluster candidates have high success rates ($> 90$\% e.g.~\citealt{Venemans2007,Kurk2004b,Maschietto2008}). In addition to selecting a relatively clean sample of \pc\ galaxies, selection based on \ha\ flux identifies star-forming galaxies that have strong continua (in comparison to Ly$\alpha$ emitters), therefore their properties can be obtained through modelling of multi-band photometry. Thus they are the ideal population with which to study the environmental dependancy of galaxies.

The high density of \pc s means large samples of \pc\ \ha\ emitters can be obtained using  instruments with relatively small fields-of-view, however, the rarer field samples require wide-field instruments. With the advent of large near-infrared imagers, a large field sample of \ha\ emitters has recently been obtained to similar depths as \pc\ samples (primarily through the HiZELS program; \citealt{Best2010}). 

In this article we constructed \ha\ and rest-frame $R-$band continuum luminosity functions of the \ha\ candidates in \pc s, and compared them to field \ha\ emitters at the same redshift. For the \pc\ environment we selected the fields surrounding the HzRGs MRC\,1138-262 at $z=2.156$ and \rg\ at $z=2.349$. Both of these radio galaxies have redshifts that place \ha\ emission from nearby galaxies within the bandpasses of available ESO/ISAAC filters.

MRC\,1138-262 lies within a well-studied \pc, with more than 30 spectroscopic redshifts (\citealt{Pentericci2000,Kurk2004a,Kurk2004b,Doherty2010}; Kuiper et al.\,sumbitted). \rg,  also known as USS\,$1707+105$, has similar properties to MRC\,1138-262. It has a stellar mass of $1.5\times10^{11}$\Msun\ \citep{Seymour2007}, clumpy optical and near-infrared morphology, and is surrounded by several nearby galaxies \citep{Pentericci1999, Pentericci2001}.  However, at the start of this project, little was known about its large-scale environment. Using narrow-band near-infrared images we searched for \ha\ emitting galaxy candidates at the same redshift as \rg.  
We compared the number densities of this population to those of the \pc\ surrounding MRC\,1138-262, and a control field, to show that \rg\ also lies in a \pc\ environment.

In Section \ref{data} we describe the observations, data reduction and source selection for the radio galaxy and control fields. In Section \ref{results} we compare the densities of \ha\ emitters in the 3 fields, and present the \ha\ and rest-frame $R$ continuum luminosity functions of each field. In Section \ref{discussion} we interpret our findings in terms of the star formation rate, stellar masses and specific star formation rates of galaxies in \pc s, and we summarise our findings in Section \ref{conclusions}.
Throughout we used the Vega magnitude scale and assumed $\Lambda$CDM cosmology, with H$_0$=70, $\Omega_m=0.3$, and $\Omega_\Lambda=0.7$. 

\section{Observations, data reduction and candidate selection}
\label{data}
\subsection{\rg\ }
\subsubsection{Observations and data reduction}
\begin{figure*}
 \begin{center}
 \includegraphics[width=2.\columnwidth]{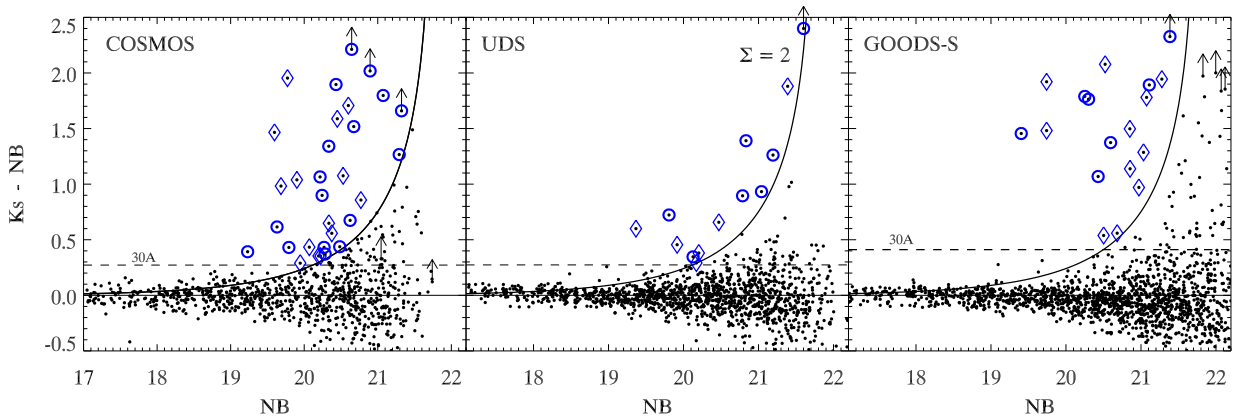}
\caption{A colour-magnitude diagram showing the selection of NB-excess objects in the COSMOS, UDS and GOODS-S fields. All sources detected in the \NB\ images are plotted as black dots. The dashed lines mark the $Ks-NB$ colour equivalent to $EW_0$ of 30\AA.  This corresponds to a slightly redder colour for the GOODS-S field than the other control fields because the GOODS-S \NB\ filter is slightly narrower than the $NB$ filter used for the other control fields.  The solid lines mark the $Ks-NB$ colour corresponding to 2 times the combined uncertainty in the \NB\ and $Ks$ images of the shallowest data (COSMOS). Candidates in all fields were selected to the same magnitude limits.  \nbexcess\ candidates, marked by blue symbols, were selected as sources with $EW_0>30$\AA\ and $\Sigma_{\rm COSMOS}>2$. The \ha\ emitters are marked by circles, whilst emission-line contaminants are marked by diamonds (see Section \ref{contaminants} for details).\label{field-selection}}
\end{center}
\end{figure*}

4C+10.48 was observed in service mode using HAWK-I \citep{Kissler-Patig2008} to obtain the $Ks$ image, and  ISAAC to obtain the $Js$ and narrow-band 2.19\micron\ (\NB\ hereafter) images. The $Ks$, $NB$ and $Js$ data were obtained in September 2008, March 2009, and April 2005 respectively. Details of the observations are given in Table \ref{tab:obs}. The radio galaxy was positioned at the centre of the ISAAC detector, and at the centre of one of HAWK-I's 4 detectors. 

The data were reduced with the ESO/MVM data reduction pipeline \citep{Vandame2004}, and underwent the usual near-infrared reduction steps of dark subtraction, flat-field removal, sky-subtraction, and creation of bad pixel and weight maps. The 4 chips of the HAWK-I images were normalised. The first-guess astrometric solutions were derived from 2MASS catalogues, but the relative astronomy was corrected with {\sc IRAF} tasks {\sc geomap}. The pixel scale of the $Ks$ HAWK-I image (0.106\arcsec /pixel) was degraded to the ISAAC pixel scale of 0.148\arcsec/pixel. 

To ensure the image depth was approximately consistent across the whole image, regions which had less than 30\% of the maximum exposure time were masked out. 
The 1$\sigma$ image depths given in Table \ref{tab:obs} were measured by placing apertures at multiple random locations. The total overlapping area of the \NB\ and $Ks$ images is 6.9\,arcmin$^2$, and the overlapping area of the $Js$, $NB$ and $Ks$ images is 5.7\,arcmin$^2$.

The $NB$ and $Js$ images were each flux calibrated using 2 standard stars taken before and after the science observations. The $Ks$ image was flux calibrated using 2MASS stars in the field of view, and further adjustments were made to this calibration by comparing the ${NB}-{Ks}$ colour of stars in the images to the predicted colours of stars in the Pickles stellar library. Uncertainties in the flux calibration are $<0.04$\,mag. No correction was applied to account for Galactic extinction as this is negligible.

Source catalogues were obtained with {\sc SExtractor} in double image mode. The \NB\ image, weighted by the square root of the exposure time was used as the detection image for selecting \NB\ excess galaxies. A similarly weighted $Ks$ image was used as the detection image to obtain matched catalogues from the $Ks$ and $Js$ images. The {\sc SExtractor} detection parameters were optimised to increase the image completeness whilst minimising spurious detections.  Sources were selected as objects with 9 connected pixels with fluxes 1$\sigma$ above the local background noise. This selection resulted in no detections in the inverted detection image, thus we do not expect spurious sources in the catalogues. Total magnitudes were taken as {\sc auto} magnitudes from SExtractor. To obtain accurate $Js-Ks$ colours, the $Ks$ image was convolved to matched the $Js$ image using the {\sc iraf} task {\sc psfmatch}. The PSF in the $NB$ and $Ks$ images were so similar that no convolution was required. $Ks-NB$ colours were measured in 1.5\arcsec\  (10 pixel) diameter apertures, whilst $Js-Ks$ colours were measured in 16 pixel diameter apertures, i.e.~three times the seeing FWHM. This ensured that 90\% of the flux from a point source lay within the colour aperture. 

\begin{table}
  \centering
  \caption{Summary of the HAWK-I and ISAAC observations. The 1$\sigma$ image depths were measured using the colour apertures (i.e. an aperture diameter of 3 times the seeing FWHM). \label{tab:obs}}
 \begin{tabular}{ccccccc}
  \hfill
 Filter & Exposure time& 1$\sigma$ depth& Seeing&Completeness  \\
 & (hours)&(mag) & (FWHM) & 95\% \\ \hline
Ks&1&23.8&0.46\arcsec &21.9\\
NB\_2.19&3.25&22.9&0.44\arcsec&20.8 \\
Js&0.4&24.25&0.78\arcsec & --\\
\hline
\end{tabular}
 
\end{table}

Image completeness was measured by simulating 1500 galaxies on the NB and Ks images using the {\sc iraf} packages {\sc gallist} and {\sc mkobject}. Objects were detected with Sextractor and the 95\% completion level for the $NB$ and $Ks$ images are 20.8 and 21.9\,mag respectively.

\subsubsection{Selection of candidate \ha\ emitters}
$\ha$ emitted at $z=2.35$ is redshifted into the \NB\ passband, therefore candidate \ha\ emitters were selected as sources with excess  \NB\ flux relative to their $Ks$ broad-band flux (known as \nbexcess\ sources). To select candidate \ha\ emitters we followed the method and criteria of \citet{Bunker1995}. The \ha\ candidates must have (i) sufficient equivalent width ($EW_{\rm 0}$) and (ii) a $Ks-NB$ colour sufficiently larger than the combined noise ($\Sigma$) of the $NB$ and $Ks$ images. 

The colour-magnitude diagram for all sources detected in the \NB\ image is shown in Fig.\,\ref{ha_select}. $\ha $ candidates were selected as sources with $EW_{\rm 0}>30$\AA\ and $\Sigma>2$. We also highlight the more-secure \ha\ candidates as those that satisfy $EW_{\rm 0}>50$\AA\ and $\Sigma>3$. 24 sources are candidate \ha\ emitters with $EW_0>30$\AA\ and $\Sigma>2$, of which 11 are selected by the stricter $EW_{\rm 0}>50$\AA\ and $\Sigma>3$ criteria.

To compare these data to the control fields and the MRC\,1138-262 \pc, we ensured that the selection criteria of \nbexcess\ galaxies are identical in all fields. The shallowest data we use are the COSMOS field data described in Section \ref{control_field_data}. The dotted line in Fig.\,\ref{ha_select} marks the location of the $\Sigma=2$ detection limit of the COSMOS data. There are 19 \nbexcess\ galaxies near \rg\ with $\Sigma_{\rm COSMOS}>2$ and $EW_{\rm 0}>30$\AA. 

\subsection{The MRC\,1138-262 \pc}
The \pc\ surrounding the radio galaxy MRC\,1138-262 is one of the best studied dense regions at $z>2$ \citep[e.g.][]{Kurk2000,Pentericci2000,Kurk2004a,Kurk2004b,Miley2006,Hatch2008,Hatch2009,Tanaka2010b}. We used the results of \citet{Kurk2004a} who identified \ha\ emitters in a 12.5\,arcmin$^{2}$ field surrounding MRC\,1138-262. The field was observed using ISAAC through both the NB 2.07\micron\ and $Ks$ filters. The total exposure time was 1.6\,h in $Ks$ and 4.8\,h in NB\_2.07\micron, and the FWHM of the seeing was 0.45\,arcsec and 0.55\,arcsec respectively. The NB\_2.07\micron\ filter has a FWHM bandwidth of 26\AA, which slightly narrower than the NB\_2.19\micron\ filter. However the transmission function of the filters differ slightly, such that the total spectral width covered by both filters is comparable.

Details on the data reduction, object detection, and selection of NB-excess sources can be found in \citet{Kurk2004a}. Candidate H$\alpha$ emitters were selected following the same procedure as described above, in particular, the $\Sigma=2$ curve for the MRC\,1138-262 data follows closely the dotted curve in Fig.\,\ref{ha_select} (see Fig.\,6 in \citealt{Kurk2004a}). To allow a robust comparison we only selected NB-excess sources with rest-frame $EW_{\rm 0}>30$\,\AA. There are 38 sources with $\Sigma>2$ and EW$>30$\,\AA, of which 3 are within the Ly$\alpha$ halo of the radio galaxy, and thus not counted as part of the overall large-scale structure. 

Spectroscopy of 9 of the brightest candidate H$\alpha$ emitters are presented in \citet{Kurk2004b}. All are confirmed to have an emission line and 3 of which are confirmed to be H$\alpha$ due to the presence of [N{\sc ii}]. The other 6 objects with just one emission line are also clustered in velocity space near the redshift of the radio galaxy, which strongly suggests that they are located in a large-scale structure associated with the radio galaxy.

\subsection{Control Fields}
\label{control_field}

\begin{table}
  \centering
  \caption{Summary of the control field data comprising the GOODS-S, COSMOS and UDS fields. Image depths were measured in apertures with diameters of 3 times the seeing of the degraded resolution images. \label{tab:field_obs}}
 \begin{tabular}{lcccccc}
Field& Filter & Exposure time& 1$\sigma$ depth& Seeing  \\
 & &(hours)&(mag) & (FWHM) \\ \hline
  GOODS-S& $Ks$&2.18&24.0 & 0.5\arcsec \\ 
&NB\_2.09\micron&14.4&23.6&0.6\arcsec \\
COSMOS&$Ks$&0.78&22.9&0.7\arcsec \\
&NB H2 2.12\micron&5.4&22.9&0.7\arcsec \\
UDS&$K$&DR8&24.5&0.8\arcsec \\
&NB H2 2.12\micron&5.3&23.2&0.6\arcsec \\
\hline
\end{tabular}
 
\end{table}

\subsubsection{NB-excess galaxies}
\label{control_field_data}
Field H$\alpha$ emitting candidates at $z\sim2.2$ were selected using 2\micron\ HAWK-I narrow-band images of the GOODS-S, UDS and COSMOS fields obtained through programs 081.A-0932(A) (see \citealt{Hayes2010} for details), and 083.A-0826(A) (P.I.~P.~Best). The image of the GOODS-S was obtained with the 2.095\micron\ HAWK-I narrow-band filter which is 191.5\AA\ wide, so captures \ha\ emitted from galaxies at $2.178<z<2.207$. The images of COSMOS and the UDS were obtained with the 2.124\micron\ narrow-band filter which is 300\AA\ wide, so they capture \ha\ emitted from galaxies at $2.213<z<2.259$. All 3 fields have a similar field-of-view of 57\,arcmin$^2$.

The data were reduced using MVM in the same manner as described above. Images taken in poor seeing were not included in the final combined dataset. The UDS was not observed in the $Ks$ by HAWK-I so the DR8 $K$ image from UKIDSS was used instead. Final exposure times, seeing full width at half maximum (FWHM), and resulting 1$\sigma$ limiting magnitudes are given in Table.\,\ref{tab:field_obs}. 
The $K/Ks-$band data were flux calibrated using the publicly available catalogues of \citet{Retzlaff2010}, Simpson et al.\, (2010 in prep.) and \citet{Ilbert2009}. The NB data were flux calibrated using standard stars, and further adjustments made after examining the $Ks-NB$ colours of stars in the fields. The resolution of the images were homogenised to match that of the lowest-resolution images, and colours were measured in apertures that were 3 times the seeing FWHM (1.8\arcsec\ for GOODS-S, 2.4\arcsec\ for the UDS, and 2.1\arcsec for COSMOS). In all 3 control fields approximately 90\% of the light from a point source lies within these apertures. 

Objects were detected using {\sc Sextractor} in double image mode, using a weighted $NB$ image as the detecting image. The same {\sc Sextractor} parameter file used for the \rg\ data was also used for the control field data to ensure the source detection was as similar as possible. \nbexcess\ galaxies were selected as objects with $EW_{\rm 0}>30$\AA\ and $\Sigma>2$ (see Fig.\,\ref{field-selection}). We ensured that the selection of candidates was identical to the \rg\ field by selecting galaxies to the same $\Sigma=2$ limit. The COSMOS data are the shallowest data therefore only sources up to the $\Sigma=2$ depth of the COSMOS data were selected. This selection resulted in 18 \nbexcess\ galaxies in GOODS-S, 13 in the UDS, and 31 in COSMOS. All candidates were checked by eye to ensure they were real sources. The total combined area of all 3 control fields is 172\,arcmin$^{2}$. 

\begin{figure}
 \begin{center}
 \includegraphics[height=1\columnwidth]{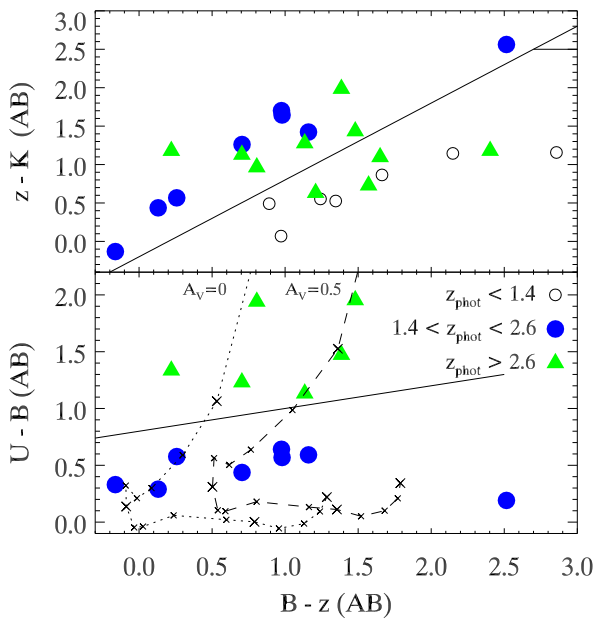}
\caption{\label{photz_selection}\ha\ emitters were selected from the sample of NB-excess objects using optical and near-infrared colours. The top panel plots the colours of all GOODS-S and COSMOS NB-excess galaxies with photometric redshifts, whilst the bottom panel plots only $BzK$ galaxies. The dotted line show the colours of a continuously star-forming galaxy, as it forms at $z=4$ and evolves to $z=0$. The dashed line shows a similar model but with $A_V=0.5$ of dust extinction. Crosses mark the colours of the galaxy at redshift intervals of $\Delta z=0.2$ from $z=0$ (bottom right) to $z=3$ (top left). Low redshift NB-excess contaminants were removed by selecting only objects that fall within the $BzK$ selection criteria. Galaxies at $z>3$ have red $U-B$ colours due to the presence of the Lyman break between the $U$ and $B$ passbands. Thus $z=2.2$ \ha\ candidates are galaxies that satisfy both the $BzK$ criterion and $U-B<0.8+0.2(B-z)$. }
\end{center}
\end{figure}

\subsubsection{Emission-line contaminants and selecting \ha\ emitters}
\label{contaminants}
Galaxies with excess flux in the 2\micron\ narrow-band may be \ha\ emitters at $z\sim2$, but they could also be emission line galaxies at low redshift (e.g.\, P$\alpha$, P$\beta$ or [Fe {\sc ii}] emitters) or [O{\sc iii}] emitters at $z\sim3$. These contaminants may be identified and removed by examining their optical colours.

To remove low redshift contaminants we selected only \nbexcess\ objects that are $BzK$ galaxies \citep{Daddi2004}, since low redshift contaminants will not lie within the BzK criteria. $BzK$ galaxies comprise both evolved ($p-BzK$) and star-forming ($s-BzK$) populations. The $z\sim2$ \ha\ emitters are likely to be star-forming $BzK$s and have similar colours to $s-BzK$ galaxies. To remove the $z=3$ [O{\sc iii}] emitters from our sample we use the Lyman break in the spectral energy distribution of galaxies. The Lyman break of $z=3$ galaxies is redshifted into the $U-$band, so  [O{\sc iii}] emitters have red $U-B$ colours or are $U-$band dropouts, whereas $z\sim2$ \ha\ emitters  have bluer colours. Thus we used the $U-B$ colour to segregate and remove [O{\sc iii}] contaminants.

The optical and near-infrared colours of the \nbexcess\ objects were obtained from the catalogues of  \citet{Santini2009}, Simpson et al.\,(2010) and \citet{Ilbert2009}. However, the  \citet{Ilbert2009} and Simpson et al. (2010) catalogues of the COSMOS and DR3 UDS fields do not contain counterparts for many \nbexcess\ objects. This difference may arise because Simpson et al.\,(2010) used the DR3 version of the UDS data to create their catalogue, whilst we use the much deeper DR8 images, and the \citet{Ilbert2009} catalogue is an $I-$band selected catalogue, so some of the faint, red $Ks$ sources may not be detected in the $I-$band. 

The colours of the objects missing from the catalogues were measured from the publicly available $U$, $B$, and $z$ images of the COSMOS and UDS fields (see  \citealt{Furusawa2008} and \citealt{Capak2007} for details). $U$ magnitudes of the UDS candidates were obtained from a 5.8\,hour CFHT image reduced by H. Hildebrandt (see \citealt{Hildebrandt2009} and \citealt{Erben2009} for details regard the reduction procedure).  The $U$, $B$ and $z$ images were homogenised to the same resolution and colours were measured in 3\,arcsec diameter apertures. The colours of galaxies in the public catalogues were compared to our measurements, and no significant differences were found.

In Fig.\,\ref{photz_selection} we show the $BzK$ and $U-B$ colours of all 25 NB-excess objects with photometric redshifts in the \citet{Santini2009} and \citet{Ilbert2009} catalogues. The $BzK$ selection efficiently removes all low redshift contaminants, and $z>3$ objects have red $U-B$ colours. Thus we separated the \ha\ emitters from the low and high redshift contaminates by selecting galaxies with the following AB colours:
\begin{equation}
(B-z)-(z-K) > -0.2 ~~~\cap~~~ U-B<0.8+0.2(B-z).
\end{equation}
32\% of the NB-excess galaxies with photometric redshifts are \ha\ emitters at $z\sim2$. But photometric redshifts are only determined for the brightest subset of candidates, so the total fraction of \nbexcess\ sources that are \ha\ emitters may be higher.

55 out of the 62 control field \nbexcess\ objects were detected in the $B$, $z$, and $K/Ks$ images, of which 33 are $BzK$ galaxies. Only 25 of the $BzK$ galaxies have $U-B<0.8+0.2(B-z)$ colours and are therefore likely to be $z\sim2$ \ha\ emitters. We estimate that only 45 per cent of the selected \nbexcess\ objects in the control fields are \ha\ emitters. This fraction is slightly higher, but still consistent given our low number statistics, with the estimates of \citet{Geach2008} and \citet{Hayes2010}.

\section{Results}
\label{results}
\subsection{\ha\ candidates near \rg}
This is the first study on the large-scale environment of \rg, so before we compare the \ha\ candidates in the radio galaxy fields and control fields, we first present the properties of the \ha\ candidates near \rg.
\subsubsection{Colours of \rg\ \ha\ candidates}
\begin{figure}
 \begin{center}
 \includegraphics[width=1\columnwidth]{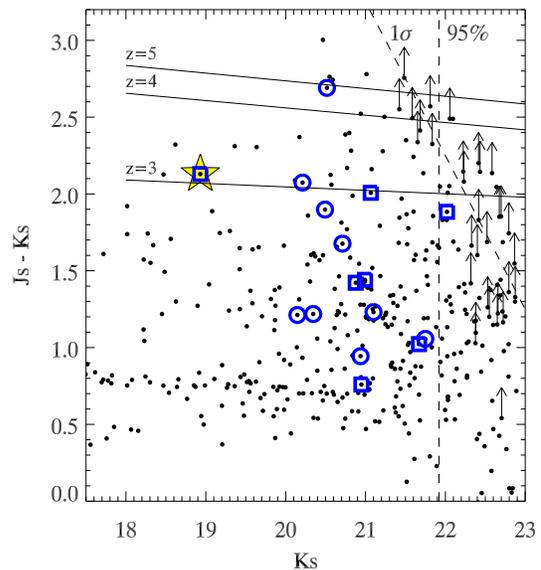}
\caption{\label{drg_select} A colour-magnitude diagram of all sources detected in the $Ks$ image of \rg. Blue symbols are \ha\ candidates (symbols defined in Fig.\,\ref{ha_select}). The yellow star marks the radio host galaxy. The expected locations of the red sequence, assuming a formation redshift of 3, 4, and 5, are plotted as solid lines. The vertical dashed line marks the 95\% completion limit of the $Ks$ image, whilst the diagonal dashed line marks the 1$\sigma$ limiting depth due to the shallow $Js$ image.}
\end{center}
\end{figure}

Fig.\,\ref{drg_select} plots the $Js-Ks$ verses $Ks$ colour-magnitude diagram for all sources detected in the $Ks$ image of \rg. The 16 \ha\ candidates detected in both $Js$ and $Ks$ images are marked by blue circles.  Unfortunately both the $Js$ and $Ks$ bandwidths cover strong emission lines from galaxies at $z=2.35$: the $Js-$band includes [O{\sc ii}], whilst the $Ks-$band includes \ha +[N{\sc ii}].  So the $Js-Ks$ colour is not a good indicator of the continuum colour for high $EW$ objects. 
 
Most of the \ha\ emitters have blue $Js-Ks$ colours in the range $0.7<Js-Ks<1.5$, but five of the \ha\ emitters lie on a possible red sequence at $Js-Ks\sim2$. One of these red \ha\ emitters is the radio galaxy. The red colour of the other two $EW_0>$50\AA\ candidates is due to the strong \ha\ emission that falls within the $Ks-$band. 

There are 3 red \ha\ candidates with relatively low $EW$s, whose red colour is likely to be due to red continuum. To cause their red colour, these galaxies must either contain a large amount of dust, or their light is dominated by an old stellar population.

\subsubsection{Spatial distribution of \ha\ emitters}
\label{distribution}
\begin{figure}
 \begin{center}
 \includegraphics[height=1\columnwidth, angle=90]{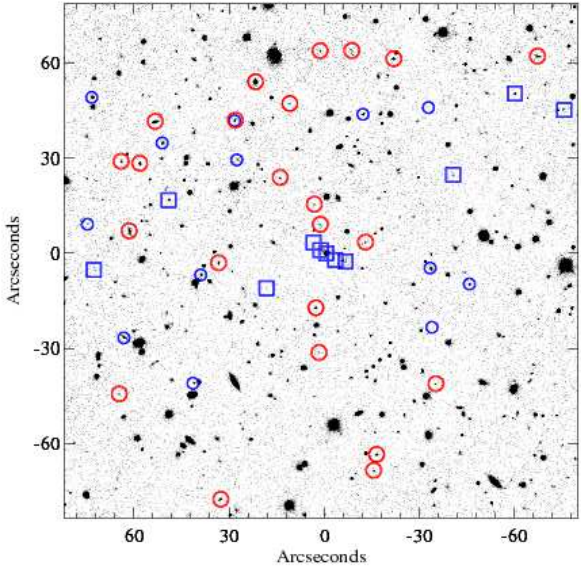}
\caption{\label{map}Spatial distribution of the  \z\,$\sim2.35$ candidates. \ha\ emitters are marked in blue: small circles denote $EW_{\rm 0}>30$\AA\ and $\Sigma>2$ candidates, whilst squares mark the more-secure $EW_{\rm 0}>$50\AA\ and $\Sigma>3$ candidates. The radio galaxy lies at the origin and distant red galaxies are marked by large red circles.}
\end{center}
\end{figure}

The spatial distribution of the \z\,$\sim2.35$ candidates around \rg\ is plotted in Fig.\,\ref{map}. The \ha\ candidates are not highly clustered, except for a strong clustering around the radio galaxy. Most of the \ha\ emitters lie in the northern half of the image, and the high $EW_{\rm 0}$ sources lie in a large East-West filament. 

Also marked by red circles are distant red galaxies (DRGs), which were selected from the $Ks$ catalogues as objects with $Js-Ks>2.3$, in accordance with the criteria of \citet{Franx2003}. These galaxies generally lie at $2<z<3.5$, so some of them may be associated with the radio galaxy.  One DRG is also selected as a \ha\ candidate which suggests it lies at the same redshift as \rg.

\subsubsection{The radio galaxy \rg}
\begin{figure*}
 \begin{center}
 \includegraphics[height=0.95\columnwidth]{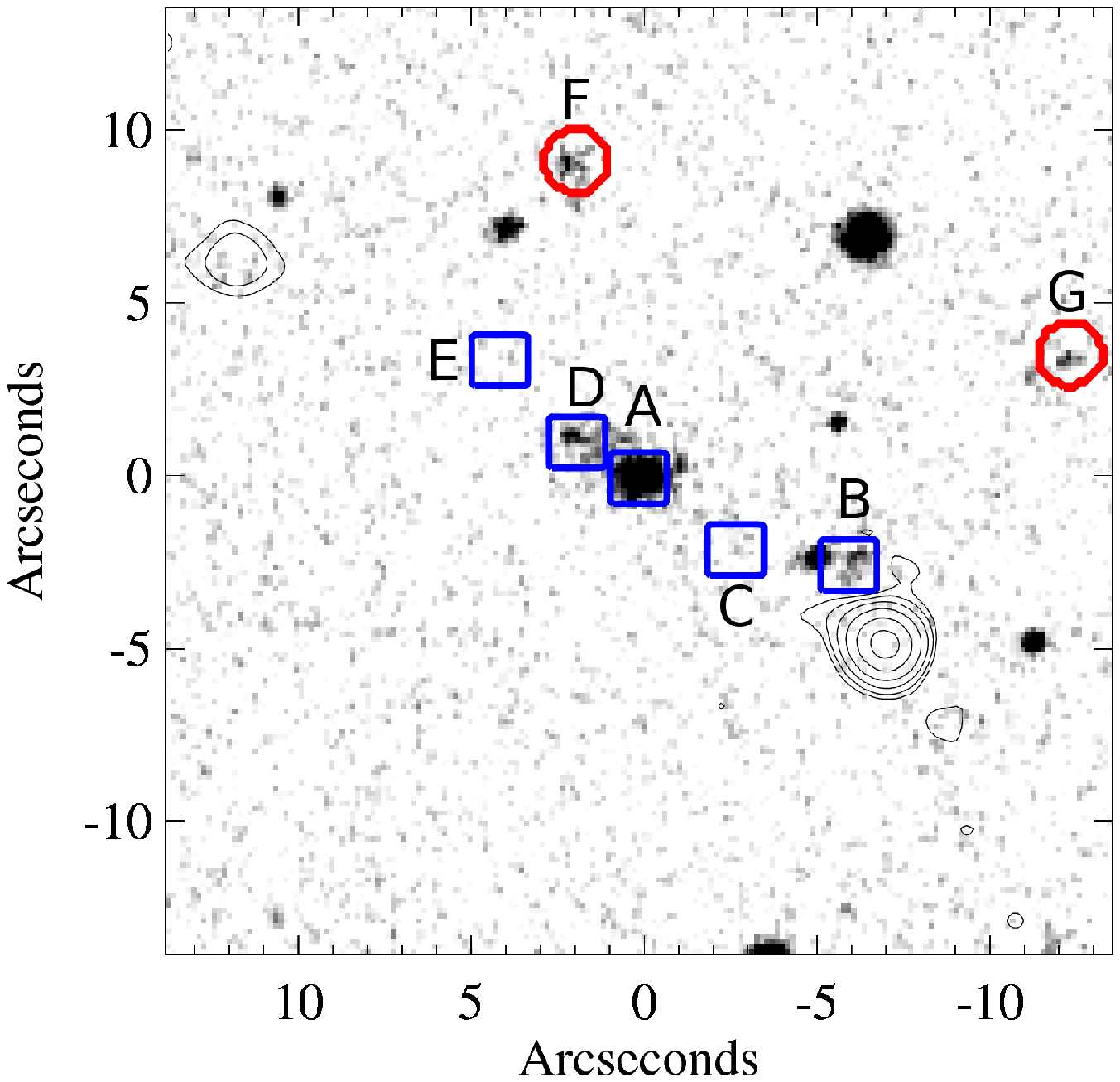}
  \includegraphics[height=0.95\columnwidth]{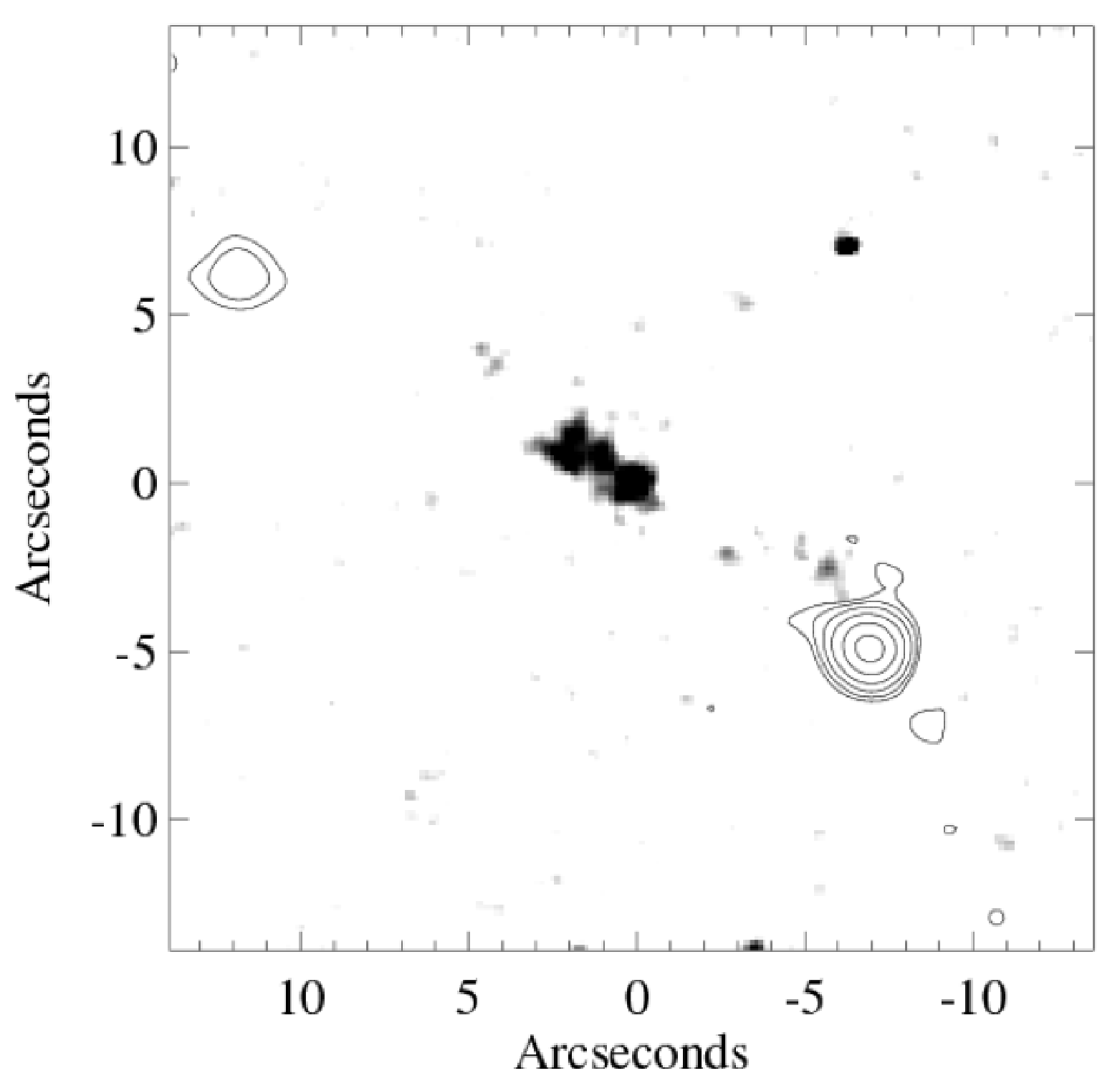}
\caption{\label{centre} Images of the central 200\,kpc around the radio galaxy \rg. The left panel is a $Ks$ image which includes both continuum and line emission. The right panel is a continuum subtracted \ha\ image. The  \z\,$\sim2.35$ candidates within 100\,kpc of the radio host galaxy (labelled A) are marked with symbols defined in Fig.\,\ref{map}. 15\,GHz radio contours are overlaid, with the outermost contour starting at 3$\times$RMS noise (0.1\,mJy) and doubling for each inner contour. The properties of the $z\sim2.35$ candidates labelled A -- G are given in Table\,\ref{tab:centre}.}
\end{center}
\end{figure*}

The distribution of \z\,$\sim2.35$ candidates within 100\,kpc of \rg\ is shown in Fig.\,\ref{centre}, and their properties are listed in Table\,\ref{tab:centre}. Within 100\,kpc of the radio galaxy there are 5 \ha\ candidates, and 2 DRGs. The DRG selection identifies galaxies across a wide redshift range so we cannot assume these DRGs have the same redshift as the radio galaxy. Never the less we show their derived masses assuming they are at $z=2.35$.
 
The stellar masses of the  \z\,$\sim2.35$ candidates were estimated from their $Ks$ magnitudes. Contamination by \ha\ emission was removed by subtracting the excess flux in the $NB$ image from the $Ks$ images. The resulting continuum magnitudes are labelled $K_{\rm cont}$. The relationship between stellar mass and $K_{\rm cont}$ for galaxies at $z=2.35$ was estimated using \citet{BC03} stellar synthesis models. Six stellar synthesis models were used, all of which assumed the \citet{Salpeter1955} initial mass function (with stellar masses between 0.1 and 100\Msun). They have a range of star formation histories, including continuous, and exponentially declining with  $\tau=1$, $0.5$, $0.1$, $0.05$ and $0.01$~Gyr. These models were redshifted to $z=2.35$ and moderate amounts of dust extinction were applied to the models (up to $A_V$=2\,mag). Finally the models were convolved with the HAWK-I $Ks$ and ISAAC $Js$ filter bandwidths to obtain the following relation:
$\log~ M_*=[K_{cont}-44.1\,(\pm 0.1)-1.4\,(\pm 0.04)\times(Js-K_{\rm cont})]/-2.5$. 

Star formation rates (SFRs) were estimated from the \ha\ luminosity of the \z\,$\sim2.35$ candidates assuming they are at $z=2.35$ and correcting for a 25\% contribution from the [N{\sc ii}] doublet in the $NB$ bandwidth. SFRs were derived using the \citet{kennicutt1998} relation appropriate for the \citet{Salpeter1955} initial mass function. This allows one to easily compare with values in the literature. To scale these SFRs and masses to the \citet{Chabrier2003} initial mass function they should be multiplied by a factor of 0.66.
 
Most of the stellar mass within 100\,kpc of the radio galaxy is concentrated in the object A, which presumably is the radio host galaxy. If we only consider the \ha\ emitters, more than 80\% of the stellar mass lies within the clump labelled A. The instantaneous star formation is more spread out, although object A still contributes 40\% of the total SFR. This central concentration of mass and widespread star formation is reminiscent of MRC\,1138-262 \citep{Hatch2009}. 

We must be cautious about the mass and SFR estimates for object A as this object may contain an active galactic nucleus (AGN) that can contribute to both the \ha\ and continuum luminosity.  However, object A is significantly extended in both the $Ks$ and $NB$ images, and there is no evidence of point-like emission from a central unobscured AGN. So it is unlikely that there is a large contribution of light from the AGN. However, a partially obscured AGN could still contaminate the $Ks$ and \ha fluxes, and our stellar mass and SFR estimate of this object could be too high. 

\subsubsection{Radio triggered star formation}

The \ha\ emitters near the radio galaxy are aligned along the radio jets at a position angle of 58$^{\circ}$. A Monte Carlo simulation of the distribution of the \ha\ emitters results in only a 0.08\% chance that the \ha\ emitters will be randomly orientated in a cone within 13$^{\circ}$ of the radio jets. Therefore their alignment with the radio jets imply a causal connection.

Radio triggered star formation has been observed in several nearby and distant radio galaxies \cite[e.g][]{vanBreugel1985,Bicknell2000}. The alignment of the \ha\ emitters with the radio jets suggest this system may be a good candidate for such jet-triggered star formation. However the emitters may also be material that is ionized by the central AGN (see \citealt{MileydeBreuck2008} for a review).

\begin{table}
  \centering
  \caption{Properties of \z\,$\sim2.35$ candidates within 100\,kpc of \rg\ which are marked and labelled in Fig.\,\ref{centre}. Objects A to E are \ha\ emitters, whilst objects F and G are DRGs. \label{tab:centre}}
 \begin{tabular}{ccccl}
  \hfill
ID & Mass  & SFR & $Js-K_{cont}$&notes \\
 & ($10^{9}$\Msun)&(\Msunpyr) &&  \\ \hline
A&141$\pm36$&127&2.0&presumed HzRG host\\
B&12$\pm2$&27&1.4&\\
C&$<3$&9&--&\\
D&10$\pm2$&120&1.4&labelled D+E in \\
&&&&\citet{Pentericci2001}\\
E&$<3$&19&--&\\
F&$35\pm10$&--& 2.5&\\
G&$32\pm10$&--&2.6&\\

\hline
\end{tabular}
 
\end{table}

In order to determine whether these are dwarf galaxies or simply clumps of ionized gas we searched for stellar continuum in other wavebands. Objects A, B and D have relatively strong $K_{\rm cont}$ continuum fluxes, likely emitted from an underlying stellar population. The stellar masses of these objects range from $10^{9}$\Msun\ to $10^{11}$\Msun. Object A and its nearest neighbour D lie at approximately the same redshift \citep{Iwamuro2003}, and are possibly interacting or merging as they are joint by a single emission line halo.  

Object C is detected in both F606W (25.6 mag) and F160W (22.1\,mag) in the Hubble Space Telescope ({\it HST}) images of  \citet{Pentericci1999, Pentericci2001}, but object E is not detected in either image.  Nebular emission (both continuum and emission lines) contributes to the flux in each of these {\it HST} bands so it is difficult to unambiguously determine whether any of the emission is stellar continuum. 

We conclude that objects A, B and D are galaxies, whose star formation may be triggered or enhanced by the radio jets. Object D is very bright in \ha, and has a similar SFR to clump A, however it contains a relatively small amount of mass. If stars were continuously forming in object D at its current rate, its stellar mass could be formed in only 65\,Myrs, which is a similar timescale to the lifetime of the radio emission. 

The nature of objects C and E is unclear, they may be dwarf galaxies with masses less than $10^{9.5}$\Msun, or simply pockets of dense gas ionized by the AGN or undergoing their first burst of star formation. However, the location of these objects suggest that the radio jets are responsible for the enhanced \ha\ emission that makes them visible. The massive radio galaxy may be surrounded by several dwarf galaxies, or pockets of {\sc Hi} gas, that cannot be detected, but may become visible when their line emission increases as they pass through the radio jets. 

\subsection{Surface density of candidate \ha\ galaxies}
In this section we compare the density of \ha\ emitters in the radio galaxy fields  to the control fields. MRC\,1138-262 lies in a dense \pc, with many spectroscopically confirmed \ha\ emitting members \citep{Kurk2004a,Kurk2004b}. If \rg\ also lies within a \pc s, it should be surrounded by a large overdensity of \ha\ emitters, of comparable density to the MRC\,1138-262 \pc .

\subsubsection{NB-excess galaxies}
\begin{figure}
 \begin{center}
 \includegraphics[width=1\columnwidth]{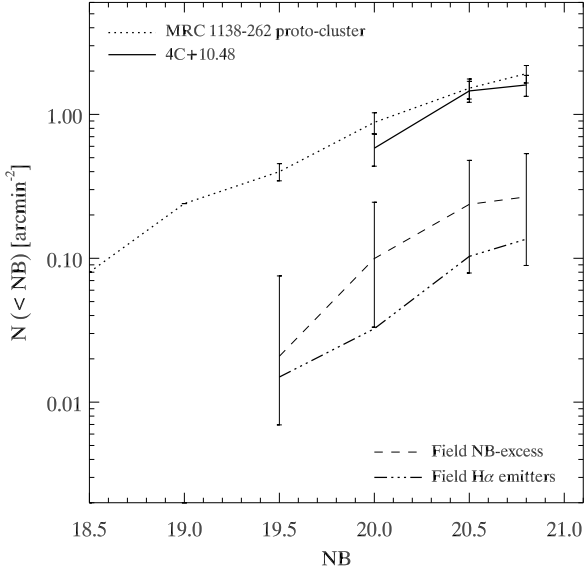}
\caption{Cumulative number counts of \nbexcess\ galaxies in the \rg\ field (solid line) and MRC\,1138-262 field (dotted line) compared to number counts of the control field (dashed line). \rg\ lies in a dense field containing 5$\pm1$ times more NB-excess objects than expected.The number density of \ha\ candidates in \rg\ is similar to the $z=2.16$ proto-cluster field surrounding MRC\,128-262, implying \rg\ resides in a similarly dense proto-cluster environment. The radio galaxies and \nbexcess\ objects within 100\,kpc of the radio galaxies are not included in the analysis as they may be influenced by the radio jets. \label{overdensity}}
\end{center}
\end{figure}
The surface density of \nbexcess\ galaxies per $NB$ magnitude is the most basic observed quantity which can determine whether a region is overdense or not. The uncertainties and significance of this measurement can be well determined from a counts-in-cells analysis of the control field. 

Five of the line-emitters in the \rg\ field are influenced by the radio galaxy (see Section \ref{distribution}), whilst 3 galaxies lie within the Ly$\alpha$ halo of MRC\,1138-262, so may also be influenced by the radio source in this field.  We are interested in the large-scale structure surrounding these radio galaxies, beyond the immediate influence of the radio jets, so these sources were removed from the catalogues and not included in the remainder of this study.

The surface densities of \ha\ candidates surrounding the two radio galaxies \rg\  and MRC\,1138-262 are shown in Fig.\,\ref{overdensity}, where they are compared to the control field. The density of \nbexcess\ objects in the GOODS-S field was adjusted to account for the difference in narrow-band filter bandwidths. 

The quoted uncertainties are 1$\sigma$ standard deviations resulting from a counts-in-cells analysis measured from 6.9 or 12.5\,arcmin$^2$ cells in the control fields. These are the sizes of the \rg\ and MRC\,1138-262 fields respectively. The 1$\sigma$ standard deviation of the density of $NB<20.8$ galaxies in a 6.9\,arcmin$^2$ cell is 0.27\,arcmin$^{-2}$, and 0.2\,arcmin$^{-2}$ for a 12.5\,arcmin$^2$ cell. 

There is a large overdensity of NB-excess objects in the radio galaxy fields, with 5$\pm1$ and $6\pm1$ times more objects surrounding \rg\ and MRC\,1138-262 respectively than the control field.  The galaxy overdensity\footnote{defined as $\delta_g=[\Sigma_{\rm observed}-\Sigma_{\rm expected} ]/ \Sigma_{\rm expected}$, where $\Sigma$ is the surface density.} of the \nbexcess\ galaxies around \rg\ is $\delta_{\rm NB} =4\pm1$, and $\delta_{\rm NB}=5\pm1$ in the MRC\,1138-262 field. The significance\footnote{Significance~$ = (\Sigma_{\rm Radio~galaxy~field}-\Sigma_{\rm control~field})/\sigma(\Sigma_{\rm control~field})$} of these overdensities are 5$\sigma$ for \rg\ and 8$\sigma$ for the MRC\,1138-262 \pc.

The highly significant overdensity around \rg\ implies this radio galaxy also resides in a dense large-scale structure that extends beyond the immediate influence of the radio galaxy, at least out to distances of 0.75\,Mpc (physical). \rg\ probably lies within a \pc, or proto-group environment of a similar density as that of MRC\,1138-262. 

\subsubsection{\ha\ candidates}
In Section \ref{contaminants} we showed that more than half of the \nbexcess\ objects selected in the control fields were not \ha\ emitters at $z\sim2$, but rather line-emitting contaminants at other redshifts. To measure the true surface overdensity of \ha\ emitters in the radio galaxy fields, the \nbexcess\ sources that are not \ha\ emitters must be removed. Unfortunately we are unable to apply the same colour selection we applied to the control field samples as we do not have the same multi-band data.

The density of \nbexcess\ objects in the control field that are {\it not} \ha\ emitters is 0.13\,arcmin$^{-2}$ ($NB<20.8$\,mag). Therefore less than 1 \nbexcess\ contaminant is expected in the 6.87\,arcmin$^{2}$ field around \rg, and less than 2 \nbexcess\ contaminants in the $12.5$ \,arcmin$^{2}$ field around MRC\,1138-262.  Whereas we observed 11 and 24 \nbexcess\ objects with $NB<20.8$\,mag near \rg\ and MRC\,1138-262 respectively. Hence less than 10\% of these objects are likely to be contaminants. We ignored this negligible fraction of contaminants and assumed that all the \nbexcess\ objects in the MRC\,1138-262 and \rg\ fields are \ha\ emitters at $z\sim2$.  

This assumption is supported by the spectroscopic data on $2$\micron\ \nbexcess\ objects near $z\sim2$ radio galaxies. \citet{Kurk2004b} showed that all spectroscopically confirmed \nbexcess\ sources near MRC\,1138-262 are likely to be H$\alpha$ emitters associated with the radio galaxy, as their velocity distribution is much more peaked than the filter response curve.  Furthermore, $2$\micron\ \nbexcess\ objects have recently been studied around the $z=2.49$ \pc\ near 4C\,23.56 \citep{Tanaka2011}. All 3 \nbexcess\ objects, for which spectra were obtained, exhibited an emission line within 230\kmps\ of 4C\,23.56, indicating they lie in a \pc\ associated with the radio galaxy.  

The H$\alpha$ emitters in the control fields were separated from the contaminants using their optical colours (see Section \ref{contaminants}). However, 7 galaxies were not detected in one or more of the $B$, $z$ or $K/Ks$ images. Three (45\%) of these  \nbexcess\ objects are likely to be \ha\ emitters, and their $NB$ flux distribution was obtained using a bootstrap analysis. We resampled 3 out of the 7 galaxies not detected in the $B$, $z$ or $K/Ks$ images 1000 times to obtain the flux distribution of the non-detected galaxies. This distribution was then combined with the flux distribution of detected galaxies.

The surface density of field \ha\ emitters is shown as the triple dot-dashed line in Fig.\,\ref{overdensity}. The overdensity of \ha\ emitters is $\delta_{\ha}=11\pm2$ (5$\sigma$)  near \rg, and $\delta_{\ha}=13\pm2$ (9$\sigma$) near MRC\,1138-262. We conclude that both radio galaxies lie within dense environments, of comparable density, and are possibly the progenitors of present day galaxy clusters or groups. 

The narrow-band filters used to detect the \ha\ candidates cover a co-moving radial distance of $50-60$\,Mpc at $z\sim2.2$, but the large-scale structures associated with the radio galaxies are likely to extend only $10-20$Mpc \citep{Hatch2010}. Therefore the surface overdensities given above are lower limits to the volume overdensity of the large-scale structures associated with the radio galaxies.

\begin{figure*}
 \begin{center}
 \includegraphics[width=2\columnwidth]{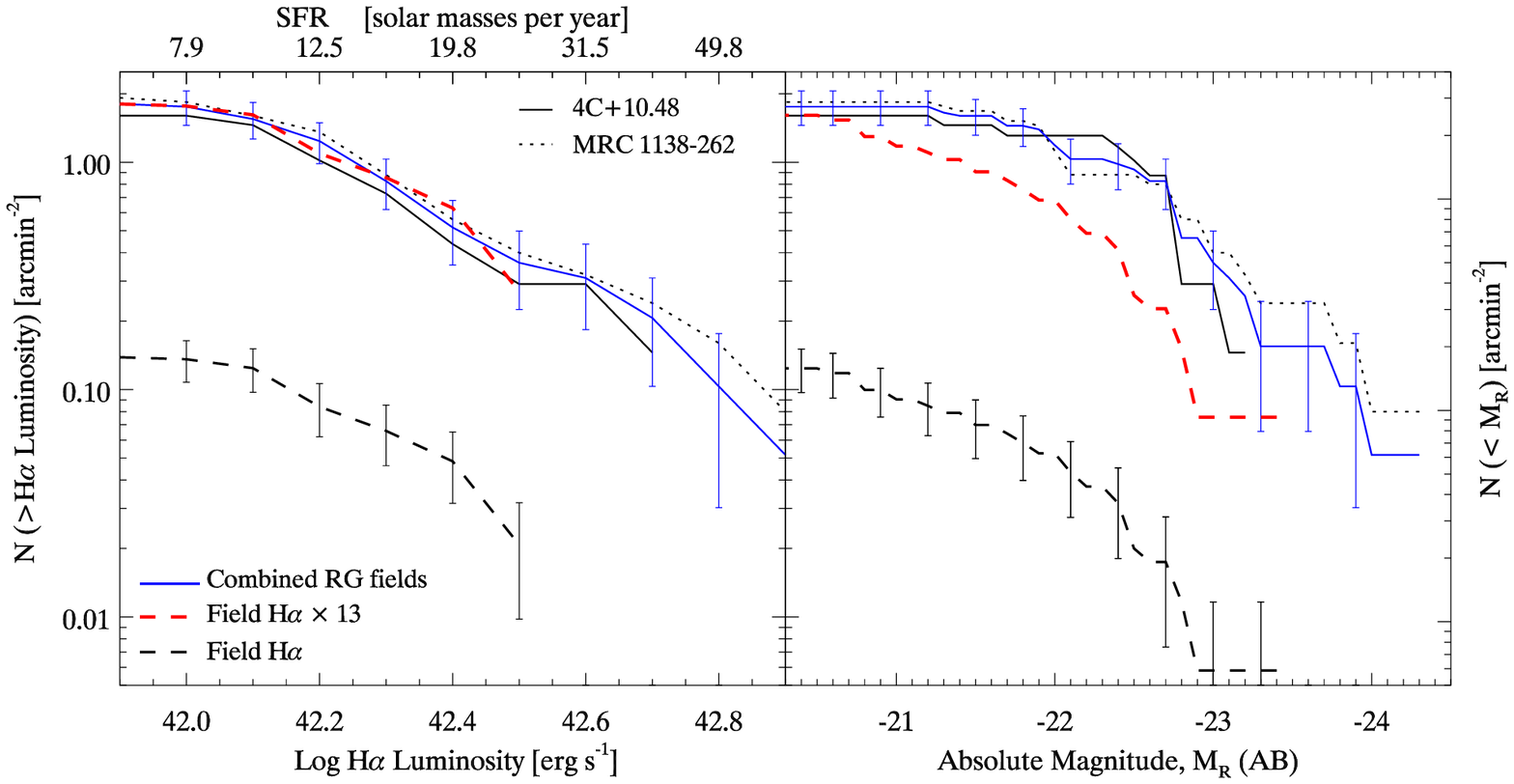}
\caption{The cumulative \ha\ LF (left) and rest-frame continuum-only $R-$band LF (right) of \ha\ galaxies in the \rg\ field (solid black lines), MRC\,1138-262 (dotted black lines) and control fields (dashed black lines). The average \pc\ LFs, resulting from the combined \rg\ and MRC\,1138-262 data, are shown as  solid blue lines. \rg\ and MRC\,1138-262 are $12$, and $14$ times denser than the field respectively. The red dashed line is the field LF re-normalised to the same density as the average \pc. The shape of the \pc\ and control field \ha\ LFs are in good agreement, whilst the \RLF s are not -- the \pc\ galaxies are typically 0.8\,mag brighter than the field galaxies. \label{LF}}
\end{center}
\end{figure*}

\subsection{Comparison of \pc\ and control field \ha\ emitters}

We have shown that the \ha\ emitters near the radio galaxies exist in an environment that is at least 10 times denser than the control fields. In this Section we examine how the star formation rate and stellar mass of the \ha\ emitting galaxies depend on environment, by comparing the \ha\ and continuum luminosity functions of emitters in the dense radio galaxy fields to emitters in the control fields. 

\subsubsection{\ha\ luminosity function}
Continuum magnitudes in the $Ks-$band, labelled $K_{\rm cont}$, were determined by subtracting the flux observed in the $NB$ image from the $Ks$ images:
 \begin{equation}
 f (K_{\rm cont})=\frac{{\rm w}_{Ks}~ f (Ks)-{\rm w}_{NB} f (NB)}{{\rm w}_{Ks}-{\rm w}_{NB}}
 \end{equation}
 where w$_{x}$ is the width of filter $x$, and $f (x)$ is the flux in filter $x$.  \ha+[N{\sc ii}]  line fluxes (abbreviated to $f (\ha)$), were then obtained by subtracting the $K_{\rm cont}$ flux from the $NB$ flux:
 \begin{equation}
f (\ha) ={\rm w_{NB} }( f(NB) - f(K_{\rm cont}))
 \end{equation}

We assumed all \nbexcess\ sources in the radio galaxy fields are \ha\ emitters since less than 10 per cent are expected to be interlopers (see discussion in Section\ref{contaminants}). However contaminants were removed from the control field sample using optical and infrared colour selection, since more than half of the \nbexcess\ sources are not \ha\ emitters at $z\sim2$. 

In order to compare the field and \pc\ samples it is essential that galaxies in the different fields are selected to the same limit. Therefore both the radio galaxy and control field catalogues were limited to galaxies brighter than $NB<20.8$\,mag, i.e., including only galaxies that are brighter than the 95\% completion limit of the shallowest $NB$ field. Sources within 100\,kpc of the radio galaxies were excluded as these may be influenced by the radio jets, however, including these galaxies does not change our results.

The left panel of Fig.\,\ref{LF} compares the cumulative \ha\ LF of \ha\ emitting candidates detected in the dense radio galaxy fields to the control field. The normalisation of the control field \ha\ LF is $12$ times below that of the \rg\ field, and $14$ times below the MRC\,1138-262 \pc, at all \ha\ luminosities. 

We combined the galaxies in the \rg\ and MRC\,1138-262 fields together to form the distribution of the \lq average' radio galaxy field (or average \pc), and plot this as a blue line. The dashed red line shows the control field \ha\ LF re-normalised to match the density of the averaged radio galaxy fields. The re-normalised LF is in complete agreement with the LF of the radio galaxy fields. A Kolmogorov-Smirnov (KS) test of the two groups results in a probability of 97\%, so there is no difference in \ha\ luminosities of galaxies in different environments at $z\sim2$. The simplest interpretation of this is that there are more star forming galaxies in the \pc s, but the dense environment does not greatly alter the \ha\ luminosities of the star forming galaxies.

\subsubsection{Rest-frame $R-$band luminosity function}
For galaxies at $z\sim2.2$ the observed $Ks$ luminosity approximately corresponds to the light within the rest-frame $R-$band, so the observed $K_{\rm cont}$ magnitudes were converted into continuum $R$ absolute magnitudes. The $R-$band luminosity functions (\RLF s) of the \ha\  emitters in the dense radio galaxy fields and the control field are shown in the right panel of Fig.\,\ref{LF}. 

The completion limit of the \RLF s is hard to define since the galaxies were selected in $NB$ not $Ks$. However the \ha\ emitters were selected in exactly the same manner from each dataset, and only galaxies up to the 95\% completion limit of the shallowest field were included. Therefore all datasets are consistent and we can robustly compare the LFs at faint magnitudes.

The density of control field galaxies with $M_{R}>-20.4$ is $12$ times below that of the density in the \rg\  field and $14$ times below the MRC\,1138-262 field. This is consistent with the overdensities derived from the \ha\ LFs.  The shape of the MRC\,1138-262 and \rg\ \RLF s are similar suggesting that the \ha\ emitters in both dense fields have similar luminosities. 

The red dashed line is the control field \RLF\ re-normalised to the average \pc\ density. By construction, the faint end of these cumulative LFs match, and the red line cannot be shifted up any further. Any discrepancy in the shapes of the LFs must therefore be due to differences in the luminosities of the galaxies in different environments.  

Although the \RLF s of the two dense radio galaxy fields are in close agreement, the shape of the control field \RLF\ does not resemble the \pc s \RLF s.  The slope of the control field \RLF\ is steeper and lies below the \pc\ LFs. A KS test of the two groups of \ha\ emitters results in a probability of 0.05. This means there is a statistically significant difference between the continuum luminosities of galaxies from the different environments.  The median difference in $R-$band luminosity between the two samples is $0.8$ magnitudes. Therefore the \ha\ emitting galaxies within the dense environment are typically 0.8\,mag brighter than their field counterparts.

These \RLF s provide further evidence that the \ha\ candidates near \rg\ are associated with the radio galaxy. The \rg\ \RLF\ matches the \pc\ \RLF\ of the MRC\,1138-262 field, but not the control field \RLF. Therefore the \ha\ emitters near \rg\ are likely to be in a large-scale structure such as a \pc.

\section{Discussion}
\label{discussion}
\subsection{SFRs and stellar masses of galaxies in dense environments}
\label{joint}

In the previous section we derived the LFs of the observed quantities. In this section we discuss the physical quantities that are derived from the observables, namely star formation rate (SFR) and stellar mass.

\begin{figure}
 \begin{center}
 \includegraphics[width=0.95\columnwidth]{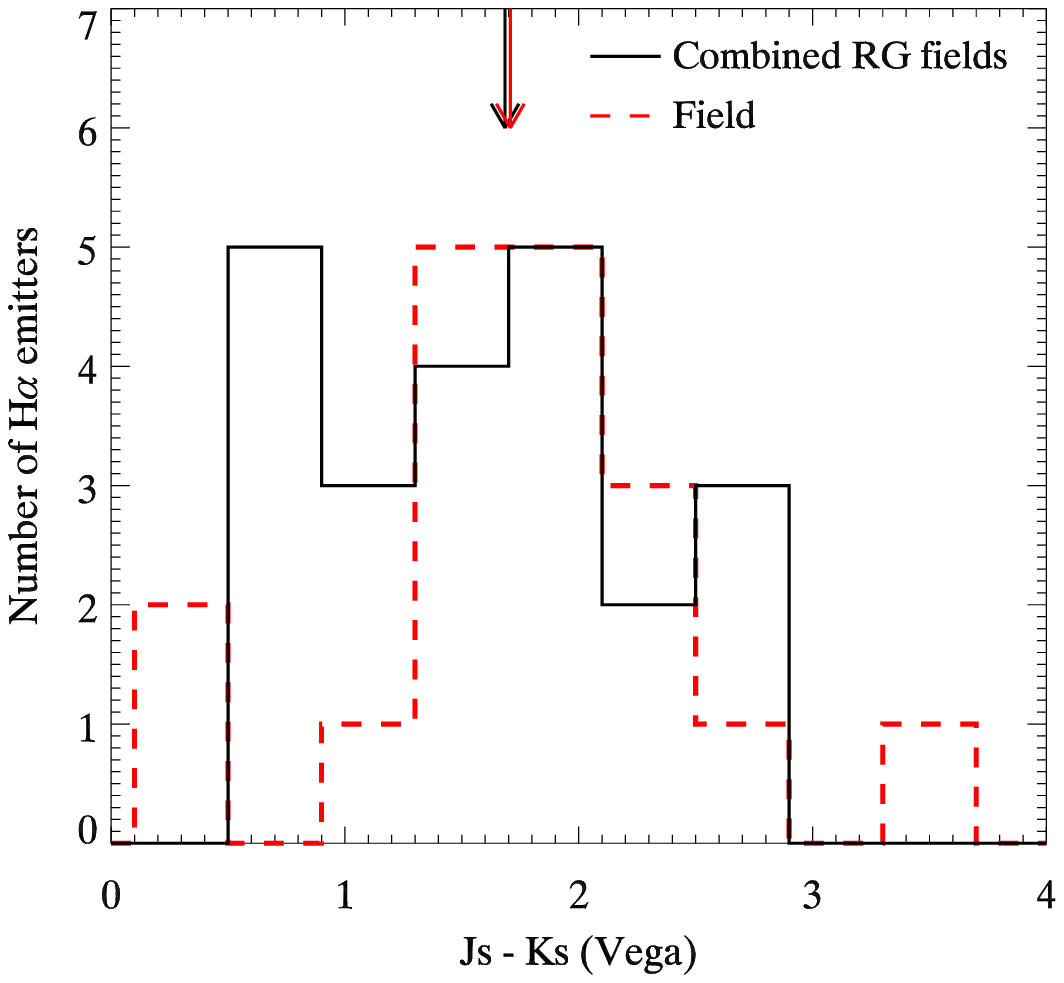}
\caption{Comparison of the observed $Js-Ks$ colours of \ha\ emitters in the dense \rg\ and MRC\,1138-262 \pc s (solid line) and those in the control fields (dashed line). A KS test of these two distributions has a probability of 67\% of being drawn from the same parent distribution, hence there is no significant difference in the colours of star forming galaxies in the field and \pc s. The median J-K colour for each population are marked by the arrows and is 1.68 and 1.71 for the \pc\ and field galaxies respectively. \label{colours}}
\end{center}
\end{figure}

\subsubsection{Star formation rates}
The \ha\ luminosities may be converted into SFRs if the \ha\ flux is due to photoionisation by young stars, and not AGN. Only 1 out of the 13 \ha\ emitters in the COSMOS field were detected in X-rays \citep{Brusa2007}, whilst only 1 out of 40 candidates near MRC\,1138-262 were detected in X-rays (not including the radio galaxy), to a similar detection limit of $10^{15}$\ergpscmps\ in the soft X-ray band \citep{Pentericci2002}. Additionally, only 1 out of 7 of the GOODS \ha\ emitters was detected in the 2\,Msec Chandra Deep Field South image \citep{Luo2010}, however, it has a soft X-ray flux slightly below the detection limit of the MRC\,1138-262 data. From these low detection rates we infer that the \ha\ emitters are  predominantly star forming galaxies, and the AGN contamination of both \pc s and field samples is likely to be low.

Fig.\,\ref{LF} shows that the normalisation of the \pc\ \ha\ LFs differ from the control field, but the shape of the LFs are consistent. Thus there are more star forming galaxies in the \pc s than the field,  but the dense environment does not alter the star formation rates of these star forming galaxies. If the SFRs of the \pc\ galaxies were lower than that of field galaxies, the slope of the \pc\ \ha\ LF would be steeper and would lie below the field \ha\ LF. Therefore the shape of the \ha\ LFs imply that the SFRs of strongly star-forming galaxies are not greatly affected by their environment.

This is similar to what is observed in low and intermediate redshift clusters \citep[e.g.][]{Balogh2002,Koyama2010}. The normalisation of the LFs differ, but the shape of the bright-end of the cluster \ha\ LFs matches the field. We did not probe the faint end of the \ha\ LF, which may depend on environment (at least at $z\sim0.8$; \citealt{Sobral2010}). So deeper narrow-band images are required to determine whether the \pc\ environment affects SFR at $z>2$.

The SFR-density relation in the distant Universe is a controversial topic as some studies find the fraction of star forming galaxies decreases with density \citep{Patel2009}, while others find that it stays constant or even increases \citep{Elbaz2007,Cooper2008,Ideue2009,Tran2010}. Unfortunately we have no measure of the non-star forming galaxy population in the \pc s, so we are unable to measure the fraction of star forming galaxies. The passive and weakly star forming populations within \pc\ remain elusive as photometric redshifts at $z>2$ are not reliable enough to determine cluster membership.

We used the difference in normalisation between the \pc s and control field \ha\ LFs to derive the SFRs for the entire \pc s. The SFR density of the field at $z=2.2$, determined from \ha\ emission, is 0.215$\pm0.09$\Msunpyr Mpc$^{-3}$ \citep{Hayes2010}, so the total SFR observed within the 6.87\,arcmin$^{2}$ field around \rg\ is $\sim 3000$\Msunpyr, and $\sim 5000$\Msunpyr\ within the 12.5\,arcmin$^{2}$ field around MRC\,1138-262. There is no observed edge to the extent of the galaxy overdensity  in these fields, and the rates do not include the radio galaxies, so these are lower limits to the total SFR in the \pc s.  These \pc s are sites of intense star formation, much more so than local clusters, whose SFRs are typically only a few hundred solar masses per year \citep{Koyama2010}.

\subsubsection{Stellar masses}

The rest-frame $R-$band luminosity may be used as a proxy for stellar mass, however, the mass-to-light ratio of galaxies depends on the age of the stellar population. In order to use the $R-$band luminosities to compare the masses of the field and \pc\ galaxies, we first checked whether the ages of the stellar populations in these samples are consistent.

For \ha\ emitters at $z\sim2.2$ the $Js-Ks$ colour is the rest-frame $U-R$ colour. This colour is a good indicator of the average age of the stellar population, (although this colour is also affected by dust reddening). Fig.\,\ref{colours} compares the colours of the \ha\ emitters within the \pc s to those in the control field. No correction was applied to account for the difference in filter passband (e.g., $Ks$ and $K$). 

The colour distributions are broad, implying \ha\ is emitted from galaxies with a wide range of stellar ages and dust contents. But the median colours of both populations are very similar ($Js-Ks\sim1.7$) and there is no significant difference in the distribution of these populations: a KS test results in a probability of 67\%. 

Since the colours of the field galaxies do not differ from the \pc\ galaxies we conclude that the mass-to-light ratios of the populations are similar, so the $R-$band luminosity can be considered a proxy for stellar mass. However the distributions do differ slightly: there are a higher fraction of blue galaxies in the \pc s than in the field sample.

We have shown that \ha\ emitters in dense environments are typically 0.8\,mag brighter than similarly selected galaxies from the control field. This luminosity difference corresponds to a factor of $\sim$2 increase in stellar mass, so the \pc\ galaxies are typically twice as massive as the control field galaxies.

We have interpreted the observed luminosity difference between the \pc\ and field galaxies as a difference in stellar mass, however, this is not the only possible interpretation. A similar luminosity offset could be observed if the \pc\ galaxies were younger and had shorter star formation timescales than the field galaxies. We do not have enough information in the limited dataset presented in this article to decisively differentiate between these two interpretations, however the similar $J-Ks$ colours implies the ages of the galaxies in both samples do not differ greatly. This suggests that the luminosity difference is likely caused by a difference in stellar mass. 

The literature provides further supporting evidence for this interpretation. \citet{Steidel2003} also found that the stellar masses of galaxies in a $z=2.3$ \pc\ are twice as massive as similarly selected galaxies in the field, and \citet{Kuiper2010} found a similar result for Lyman break galaxies in a $z=3.1$ \pc. Furthermore, \citet{Tanaka2010b} found the ages of galaxies in the MRC\,1138-262 \pc\ were similar to those in the field. Thus is it likely that the luminosity difference observed in the \RLF\ is caused by a systematic difference in stellar mass between \pc\ and field galaxies.

In the nearby Universe, both the galaxy mass and luminosity functions change with environment. The characteristic mass and luminosity increase with density, such that clusters typically contain a higher fraction of massive galaxies than the field \citep[e.g.][]{dePropris2003,Croton2005,Baldry2006}. This segregation of stellar mass with large-scale environment is seen up to $z\sim1.8$ \citep{Salimbeni2009,Scodeggio2009}, and we have now shown that this segregation persists in the early Universe, at $z>2$. 

Our results suggest that the mass function of cluster galaxies differed from field galaxies, even  before the cluster virialized. There is a well-documented relation between galaxy mass and other properties, such as star formation activity and morphology. Therefore the difference in galaxy properties between cluster and field environments may evolve from this difference in mass function at early times.

\subsection{Specific star formation rates of \pc\ galaxies}

\begin{figure}
 \begin{center}
 \includegraphics[width=0.95\columnwidth]{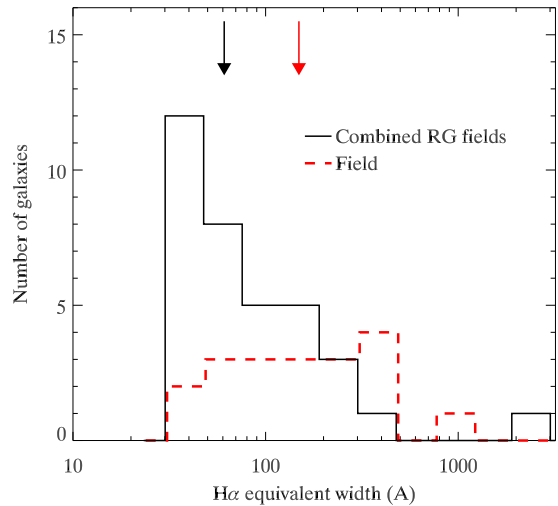}
\caption{The rest-frame \ha\ equivalent width ($EW_{\rm 0}$) of galaxies in the radio galaxy fields (\rg\ and MRC\,1138-262) is shown in black and compared to the $EW_{\rm 0}$ distribution of field \ha\ emitters. The median $EW_{\rm 0}$ of each population are marked by arrows and show that the median $EW_{\rm 0}$ of galaxies in the \pc s is 2.5 times lower than the median $EW_{\rm 0}$ of galaxies in the field.  The Mann-Whitney U-test results in a probability of 0.01 that these populations are drawn from the same parent distribution, so the $EW$s of these populations differ significantly.  \label{EW}}
\end{center}
\end{figure}

The \pc\ and control field galaxies have similar SFRs, but the \pc\ galaxies are more massive. Hence the \pc\ galaxies have lower specific star formation rates (SSFRs -- SFR per unit stellar mass) than the field galaxies.

A complementary method to examine the SSFRs of the \ha\ emitters is to study their rest-frame \ha\ equivalent widths ($EW_{\rm 0}$). The $EW_{\rm 0}$ is the ratio of the \ha\ line strength (a measure of the SFR of the galaxy) to the underlying continuum (a measure of the stellar light and mass). Hence the $EW_{\rm 0}$ of the \ha\ emission line measures the SSFR of the galaxy. Since the LFs in Fig.\,\ref{LF} suggest that the SSFRs of the \pc\ and field galaxies differ, we expect a significant difference between the distribution of $EW$s from these two populations.

Fig.\,\ref{EW} compares the $EW$s of galaxies in the 2 radio galaxy field,  to the \ha\ emitters in the control fields. Only galaxies with  $NB$ magnitudes brighter than the 95\% completeness limit of the shallowest data are included in the sample. 

The \pc\ galaxies typically have lower $EW$s than the field galaxies. The median $EW_{\rm 0}$ of the field population is 150\AA, whilst the median $EW_{\rm 0}$ of the \pc\ population is only 60\AA. Comparing the $EW$s of these populations with the KS test results in a probability of 0.13, which is low, but not significant. The Mann-Whitney U-test, which compares the difference in the mean of the two populations results in a probability of 0.01. Thus these populations have statistically different means, and are therefore not likely to be drawn from the same parent distribution. Therefore we confirm our finding that  \pc\ galaxies have lower SSFRs than their counterparts in the field.

The lower SSFRs of the \pc\ galaxies imply they formed more of their stars earlier than field galaxies. A similar result was found by \citet{Tanaka2010b}, who performed spectral energy distribution fitting of \pc\ candidates near MRC\,1138-262. They found that the \pc\ galaxies have shorter star formation timescales than field galaxies, but similar ages. Thus they also found that the MRC\,1138-262 \pc\ galaxies formed most of their stars earlier than field galaxies. These results suggest that galaxy growth is accelerated in \pc s relative to the field, as predicted by \citet{Gunn1972}.
 
\subsection{Measuring the galaxy overdensity of \pc s}
Three different measures of the overdensity of star forming galaxies near the radio galaxies have been presented: (i) the overdensity of \ha\ emitters per unit $NB$ magnitude, (ii) the overdensity of SFR determined by the relative normalisations of the \ha\ LFs,  (iii) the overdensity of \ha\ emitters per unit rest-frame $R-$band (continuum) magnitude. All three measures show that there is a larger number of star forming galaxies near the radio galaxies than in the control field. To estimate the mass and collapse time of the \pc\ we need the galaxy and mass overdensity of the \pc. Do any of these measurements provide us with the true galaxy overdensity of a \pc? 

Since we do not know whether the fraction of star forming galaxies increases or decreases in dense environments at $z\sim2$, we cannot use the normalisation of the \ha\ LF relative to the field to obtain the galaxy overdensity of the \pc. This normalisation only tells us the overdensity of star-forming galaxies in the \pc, not about the total number of galaxies in the \pc. For example, using the density of \ha\ emitters to measure the galaxy overdensity of a local cluster would underestimate the true galaxy overdensity because star formation is suppressed in local clusters. 

The continuum LF in \pc s (and the stellar mass function by inference) differs from the field. So the galaxy overdensity cannot be determined by the excess number of galaxies up to a magnitude limit, as the overdensity will depend on the limit used. For example, there are 12 times more emitters with $M_{\rm R}<-20.5$ near \rg\ compared to the field, but 25 times more emitters if we only consider galaxies with $M_{\rm R}<-22$. 

Finally, since the $NB$ magnitude of the \ha\ emitter is a combination of the continuum and \ha\ flux, the density of galaxies per $NB$ magnitude is affected by both of the above issues. The  overdensity per $NB$ magnitude is therefore hard to interpret in terms of the physical overdensity of galaxies. 

To obtain the true galaxy overdensity we must measure the galaxy stellar mass function of all galaxies (passive and star forming) in a \pc, and compare it to the field.  This provides a significant observational challenge as precise photometric redshifts at $z>1.5$ are difficult to achieve, and many \pc\ galaxies are too faint to obtain a spectroscopic redshift in a reasonable observing time.

\section{Summary and conclusions}
\label{conclusions}
We compared $z\sim2$ star forming galaxies in 2 dense \pc s to similarly-selected field galaxies. The \pc\ galaxies were selected from the catalogue of candidate \ha\ emitters from the well-studied $z=2.16$ \pc\ near MRC\,1138-262, and is supplemented with \ha\ candidates surrounding the massive $z=2.35$ radio galaxy \rg. 

We showed that \rg\ is situated in a dense environment, with $12\pm2$ times more \ha\ emitters than in the average $z\sim2.2$ field ($5\sigma$ significance). This is similar to the density of \ha\ emitters in the MRC\,1138-262 \pc, which has $14\pm2$ times more \ha\ emitter than in the field  ($9\sigma$). Due to the similar galaxy densities, we suggest that the structure surrounding \rg\ may also evolve into a group or cluster by the present day. 

A field sample of $z\sim2.2$ \ha\ emitters was constructed from 3 separate fields covering a total area of 172\,arcmin$^{2}$. Special care was taken to ensure the field and \pc\ \ha\ emitters were selected in the same fashion, so they could be robustly compared. Other line contaminants (e.g. [O{\sc iii}] emitters at $z\sim3.3$) were removed from the field sample using optical and NIR colours. Based on the density of contaminants in the control fields, the contamination fraction of \ha\ candidates in the radio galaxy fields was expected to be negligible ($1-2$ objects in each field), so no correction was made to remove foreground or background objects in the radio galaxy fields. 

We constructed \ha\ and continuum $R$ luminosity functions of the \ha\ emitters in both environments. From a comparison of the two samples we conclude:
\begin{itemize}
\item[1.] The normalisation of the \ha\ luminosity in these \pc s is $\sim13$ times greater than the field. Thus the total star formation rate within the central 1.5\,Mpc of these \pc s exceeds  $3000$\Msunpyr. This is an order of magnitude more than observed in clusters today. 

\item[2.] Although the normalisation of the \ha\ luminosity functions differ, the shape of the field and \pc\ \ha\ luminosity functions are very similar (KS probability of 0.97). We infer this to mean the star formation rate of star forming galaxies does not greatly depend on their environment.

\item[3.] The star forming \pc\ galaxies are typically 0.8\,mag brighter in rest-frame $R$ than field star forming galaxies. We showed that the $Js-Ks$ (rest-frame $U-R$) colours of field and \pc\ galaxies are consistent, so we used luminosity as a proxy for stellar mass to infer that the \pc\ galaxies are typically a factor of 2 more massive than their field counterparts. 
Since many galaxy properties correlate with stellar mass, the mass segregation present in embryonic clusters destines the future cluster galaxies to differ from field galaxies. 

\item[4.] The specific star formation rates of \pc\ galaxies are lower than field galaxies, so \pc\ galaxies form more of their stars earlier than field galaxies. This suggests that, in the early Universe, galaxy growth proceeds at a faster rate in dense environments.
\end{itemize}

The differences observed between cluster and field galaxies today may not only be driven by environmental nurture effects. They may also arise because the cluster galaxies differed from field galaxies even before the cluster formed, when the Universe was less than a quarter of its current age. Nature, as well as nurture, plays a role in the construction of cluster galaxies.

\section{Acknowledgments}
We thank Roderik Overzier for providing comments on an early draft of the article, Will Hartley for providing us with the DR8 UDS data and catalogues, and Hendrik Hildebrandt for the UDS $U-$band image. We thank the referee for providing helpful suggestions that improved the paper. NAH acknowledges support from STFC and the University of Nottingham Anne McLaren Fellowship. JK thanks the DFG for support via the German-Israeli Project Cooperation grant STE1869/1-1.GE625/15-1.This work is based on observations with ESO Telescopes at the Paranal Observatories under programme IDs: 0.81.A-0932, 082.A-0330 and 0.83.A-0826.

\bibliographystyle{mn2e}\bibliography{TX1707_pc_v6,mn-jour}
\label{lastpage}
\clearpage
\end{document}